\documentclass[12pt]{iopart}

\usepackage{iopams} 
\usepackage{graphicx}
\usepackage{epstopdf}  

\newcommand{\s}{$^1\mathrm{S}_0$}
\newcommand{\srs}{Sr(5s$n$s\,$^1\mathrm{S}_0$)}

\begin{document}

\title[Many-body Physics with Alkaline-Earth Rydberg lattices]{Many-body Physics with Alkaline-Earth Rydberg lattices}

\author{R Mukherjee$^1$, J Millen$^2$, R  Nath$^1$ , M P A Jones$^2$ and T Pohl$^1$}
\address{$^1$ Max Planck Institute for the Physics of Complex Systems, N\"{o}thnitzer Strasse 38, 01187 Dresden, Germany}
\address{$^2$ Department of Physics, Durham University, Durham DH1 3LE, United Kingdom}

\begin{abstract}
We explore the prospects for confining alkaline-earth Rydberg atoms in an optical lattice via optical dressing of the secondary core valence electron. Focussing on the particular case of strontium, we identify experimentally accessible magic wavelengths for simultaneous trapping of ground and Rydberg states. A detailed analysis of relevant loss mechanisms shows that the overall lifetime of such a system is limited only by the spontaneous decay of the Rydberg state, and is not significantly affected by photoionization or autoionization. The van der Waals $C_6$ coefficients for the \srs~Rydberg series are calculated, and we find that the interactions are attractive. Finally we show that the combination of magic-wavelength lattices and attractive interactions could be exploited to generate many-body Greenberger-Horne-Zeilinger (GHZ) states. 
\end{abstract}


\section{Introduction}

The physics of strongly-correlated quantum systems is an important topic that cuts across the fields of condensed matter physics and quantum information. Combining high-lying Rydberg states with cold atomic gases provides an appealing approach, as the strong long range van der Waals interactions between Rydberg atoms cause strongly correlated, many-body quantum states to emerge {\it directly} as Rydberg atoms are excited in dense clouds of cold atoms. The strong interactions inhibit the excitation of neighbouring atoms to the Rydberg state in an effect known as the dipole blockade \cite{Jaksch2000,Lukin2001}. This effect has been elegantly demonstrated in experiments using two independently trapped atoms \cite{Urban2009,Gaetan2009}, where it has also been exploited to entangle qubits \cite{Wilk2010}, and perform gate operations \cite{Isenhower2010,Zhang2010}. The utility of tunable Rydberg-Rydberg interactions is not limited to two atoms, and has lead to the theoretical development of a remarkably versatile toolbox for quantum information processing  \cite{Saffman2010}.
Recent theoretical work has studied the excitation dynamics and many-body phase diagram of large Rydberg atom chains and lattices \cite{Sun2008,Olmos2009,Pohl2010,Schachenmayer2010,Weimer2010,Lesanovsky2011,Zuo2010,Tezak2010}, and demonstrated their utility for digital quantum simulation of exotic many-body Hamiltonians \cite{Mueller2009,Weimer2010b}.

A pre-requisite for many of these applications is the ability to trap Rydberg atoms and ground state atoms in a common lattice, in order to manipulate their properties without significant loss and heating. Optical trapping of alkali Rydberg atoms due to the ponderomotive force exerted by spatially modulated light fields has been studied theoretically  \cite{Dutta2000, Saffman2005,Knuffman2007,Younge2010a} and experimentally \cite{Younge2010b}. Static electric and magnetic fields can also be used to trap Rydberg atoms \cite{Hogan2008,Lesanovsky2005,Hezel2006,Mayle2009,Pohl2009}. Owing to the exaggerated response of Rydberg atoms to external perturbations, achieving identical trapping potentials for Rydberg and ground state atoms appears, however, challenging. In this respect, atoms with two valence electrons offer a promising approach, as the remaining valence electron of a singly excited Rydberg state provides an additional degree of freedom to probe and manipulate the atom \cite{Millen2010,Millen2011}.

Here we use the polarizability of the extra valence electron to realize a magic-wavelength optical lattice for ground state atoms and strongly interacting Rydberg atoms. We calculate the van der Waals coefficients for the \srs~Rydberg series, and show that these interactions are attractive. For the bosonic isotopes where the nuclear spin $I=0$, the \s~states have no degeneracy, and are therefore particularly attractive for high-fidelity quantum state preparation. We point out qualitative differences between the resulting many-body excitation dynamics and that of the previously studied case of repulsive Rydberg interactions \cite{Pohl2010,Schachenmayer2010,Weimer2010,Lesanovsky2011}, and show how the combination of attractive interactions and optical trapping can be used to prepare many-body GHZ states. This class of entangled states is a resource for entanglement-enhanced measurements \cite{Giovannetti2004}, and their creation in a lattice of strontium atoms could have applications in high-precision frequency metrology \cite{Blatt2008}.

\section{Optical traps for alkaline-earth Rydberg atoms}
\begin{figure}[t!]
\centering \resizebox{0.75\columnwidth}{!}{\includegraphics{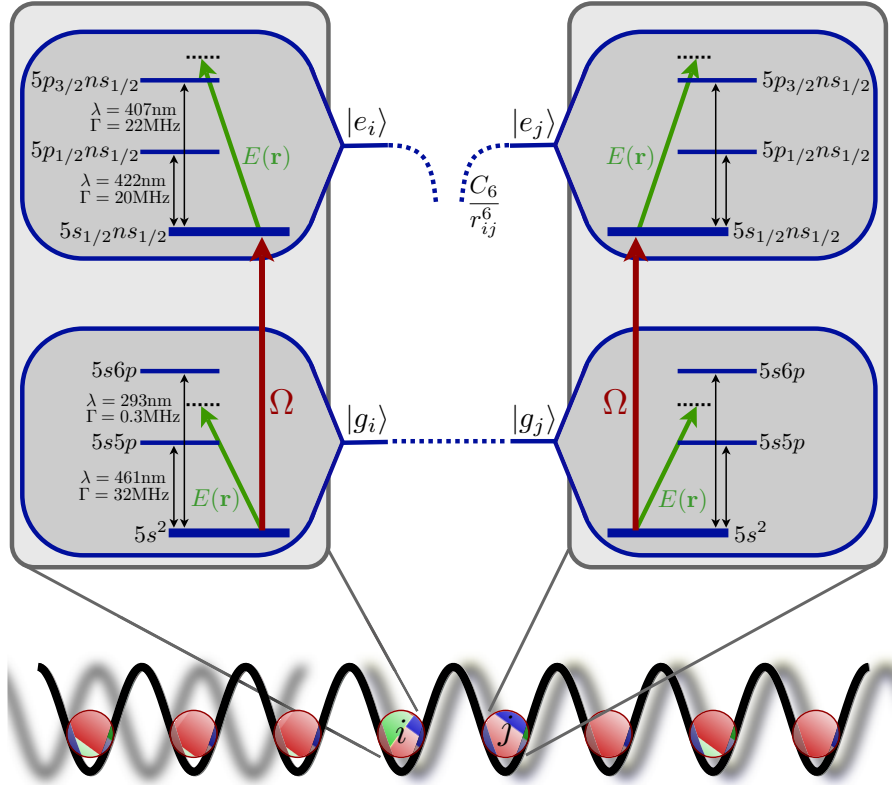}}
\caption{Level scheme for optical dressing of the Sr(5s$^2$\,$^1\mathrm{S}_0$) ground state and the Sr($n$s5s\,$^1\mathrm{S}_0$) Rydberg state to provide an optical lattice for both states. The resulting dressed ground  ($|g\rangle$) and Rydberg ($|e\rangle$) states are coupled with a Rabi frequency $\Omega$ which gives rise to strongly correlated excitation dynamics due to the strong van der Waals interactions between atoms in state $|e\rangle$.\label{fig1}}
\end{figure}

Figure \ref{fig1} illustrates the setup for forming the proposed alkaline-earth Rydberg atom lattice. Here we consider the particular case of strontium atoms, but the general discussion can be applied to the other alkaline-earth elements, or atoms with a similar electronic structure such as ytterbium. The optical lattice potential for atoms in the 5s$^2$\,\s~ground state is provided by a periodic light field with electric field amplitude $E({\bf r})$, that induces off-resonant coupling to low-lying excited states. Additional lasers drive a near-resonant two-photon transition to a singly-excited $n$s5s\,\s~Rydberg state, with a two-photon Rabi frequency $\Omega$. The remaining  5s electron allows  optical dressing of the Rydberg state via the strong (5s$n$s$\leftrightarrow$5p$n$s) transitions, and the standing wave $E({\bf r})$ therefore  also creates an optical lattice potential for the Rydberg atoms. Upon adiabatic elimination of the weakly admixed $p$-states, this setting yields effective two-level atoms composed of a dressed ground ($|g\rangle$) and excited ($|e\rangle$) state, where only the latter features strong van der Waals interactions (cf. figure \ref{fig1}). In the following, we consider the optical lattice potential for the ground and Rydberg states in more detail, and show that there are ``magic wavelengths'' where state-independent optical trapping  can be achieved. 

\subsection{Magic wavelength for ground and Rydberg states}
We consider a three-dimensional optical lattice created by three overlapping standing waves with frequency $\omega_{\rm L}$ and lattice spacing $a$. 
Because of the enormous range of the Rydberg-Rydberg interactions, we will consider large lattice spacings on the order of several $\mu$m, and, thus, considerably larger than the optical wavelength $\lambda_{\rm L}$. As demonstrated in \cite{Nelson2007}, the lattice spacing $a=\lambda_{\rm L}/[2\sin(\theta/2)]$ can be varied by adjusting the angle $\theta$ between the two co-propagating lattice beams that form each of the three standing wave fields. 
For atoms in state $|a\rangle$, optical dressing on transitions to nearby states $|b\rangle$, results in an optical lattice potential 
\begin{equation}\label{eq1}
U_a({\bf r})=\frac{\alpha_a}{2}E^2({\bf r})
\end{equation}
where
\begin{equation}\label{eq2}
\alpha_a=\sum_b\alpha^b_a=\sum_b \frac{\wp^2_{ab}}{\hbar}\frac{\omega_{ab}}{\omega_{\rm L}^2-\omega_{ab}^2}
\end{equation}
is the scalar polarizability of the atomic state $|a\rangle$, and $\alpha^b_a$ denotes the constituent polarizabilities arising from optical dressing on the respective transitions, with $\wp_{ab}$ and $\omega_{ab}$ denoting the corresponding dipole matrix element and transition frequency. 
\begin{figure}[t!]
\centering \resizebox{0.75\columnwidth}{!}{\includegraphics{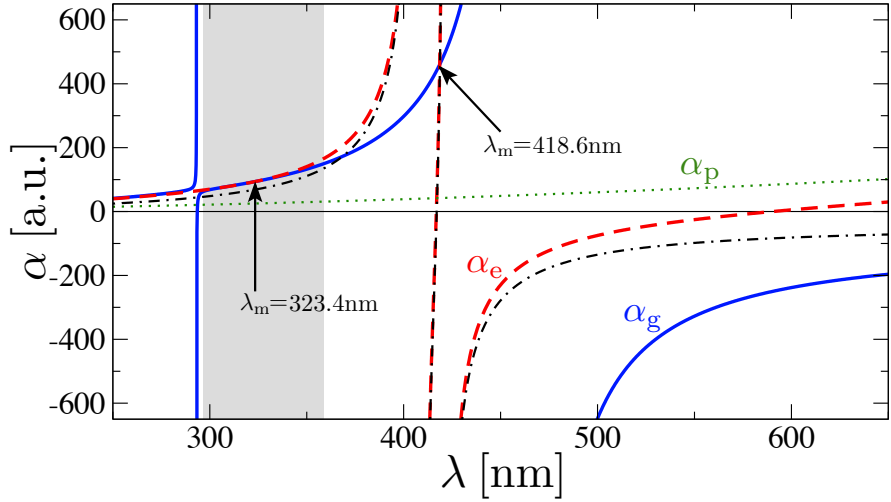}}
\caption{Scalar polarizability as a function of  wavelength. Shown are the polarizability of Sr($5s^2$) ground state atoms (blue solid line), the total polarizability of the Sr(5s$n$s)\,\s~Rydberg states (red dashed line) and the individual contributions of the Rydberg electron (green dotted line), and the core (black dash-dotted line). The two magic wavelengths at which the ground state and Rydberg state lattice potentials coincide are marked by the arrows.\label{fig2}}
\end{figure}

In addition to polarizing the deeply bound core electrons, the optical lattice field also affects the orbit of the weakly bound Rydberg electron. As shown in \cite{Dutta2000} the resulting energy shift can be accurately described within a semiclassical picture based on the ponderomotive potential $e^2 E^2/4m_{\rm e} \omega^2_L$ of a free electron with mass $m_{\rm e}$. For energetically isolated $n$s Rydberg states, the resulting atomic potential \cite{Dutta2000}
\begin{equation}\label{eq3}
U_{\rm p}({\bf r})=\int d{\bf r}^{\prime} \frac{e^2 E({\bf r}+{\bf r}^{\prime})^2}{4m_{\rm e} \omega^2_L}|\psi({\bf r}^{\prime})|^2
\end{equation}
is well approximated by the electronic ponderomotive shift averaged over the corresponding Rydberg wavefunction $\psi$. For sufficiently tight atomic confinement at the local lattice sites the Rydberg electron only probes the harmonic part of the periodic intensity pattern, such that the integral can be readily evaluated and rewritten in the form of (\ref{eq1})
\begin{equation}\label{eq3b}
U_{\rm p}({\bf r})=\frac{\alpha_{\rm p}}{2}E^2({\bf r})+\int d{\bf r}^{\prime} \frac{e^2 E({\bf r}^{\prime})^2}{4m_{\rm e} \omega^2_L}|\psi({\bf r}^{\prime})|^2
\end{equation}
with the corresponding polarizability given by the free-electron value 
\begin{equation}\label{eq4}
\alpha_{\rm p}=\frac{e^2}{2m_{\rm e}\omega_{\rm L}^2}\;.
\end{equation}

The total polarizabilities of the ground  ($\alpha_{\rm g}$) and  Rydberg  ($\alpha_{\rm e}$) states are shown in figure \ref{fig2}. Over the plotted range of optical wavelengths, one needs to consider two transition resonances, $5\mathrm{s}^2\rightarrow 5\mathrm{s}5\mathrm{p}$ and $5\mathrm{s}^2\rightarrow5\mathrm{s}6\mathrm{p}$, for the dressed ground state and two core resonances, $5\mathrm{s}_{1/2}n\mathrm{s}_{1/2}\rightarrow 5\mathrm{p}_{1/2}n\mathrm{s}_{1/2}$ and $5\mathrm{s}_{1/2}n\mathrm{s}_{1/2}\rightarrow 5\mathrm{p}_{3/2}n\mathrm{s}_{1/2}$, for the dressed Rydberg state (cf. figure \ref{fig1}). The corresponding dipole matrix elements were calculated from the atomic \cite{fuhr2005} and ionic \cite{jiang2009} transition rates. Figure \ref{fig2} also shows the individual contributions of the core resonances and the ponderomotive shift. In the alkalis, only the ponderomotive contribution is available for Rydberg atoms, leading to relatively weak lattice potentials that are always repulsive. Here, the additional core potential leads to much deeper Rydberg atom lattice potentials that can be either attractive or repulsive. Attractive (red-detuned) potentials are appealing as they are simple to realize experimentally. For strontium, red-detuned traps for Rydberg atoms are possible in the wavelength range from $\sim430$~nm to $\sim550$~nm. 

By equating the ground state polarizability $\alpha_{\rm g}=\alpha_{5s^2}^{5s5p}+\alpha_{5s^2}^{5s6p}$ and Rydberg state polarizability $\alpha_{\rm e}=\alpha_{5s_{1/2}ns_{1/2}}^{5p_{1/2}ns_{1/2}}+\alpha_{5s_{1/2}ns_{1/2}}^{5p_{3/2}ns_{1/2}}+\alpha_{\rm p}$ one obtains several magic wavelengths that facilitate trapping of ground state and Rydberg state trapping in identical optical lattice potentials. Most relevant from a practical point of view, we find one magic wavelength at 
\begin{equation}\label{eq5}
\lambda_{\rm m}=323.4\ {\rm nm}
\end{equation}
with $\alpha_{\rm m}=93.9$~a.u. and one at
\begin{equation}\label{eq6}
\lambda_{\rm m}=418.6\ {\rm nm}
\end{equation}
with $\alpha_{\rm m}=460.3$~a.u.. 
Around the former, we find a broad range of wavelengths for which the ground and Rydberg state polarizabilities are nearly identical. This range is marked by the grey area in figure \ref{fig2}, and contains the third harmonic of the  Nd:YAG laser at $\lambda=355$~nm, where single-frequency CW lasers are available with reasonable output power. At this wavelength the potentials for the two states only differ by $4\%$. The magic wavelength at  418.6~nm could be reached using commercially available tunable frequency-doubled diode laser systems. 

A similar analysis \cite{Saffman2005} for rubidium Rydberg states also revealed a magic wavelength, based solely on the balance of the atomic ground state polarizability and the ponderomotive Rydberg atom potential. As pointed out in \cite{Saffman2005}, the corresponding laser frequency lies close ($\sim1$\ GHz) to the atomic $5\mathrm{s}\rightarrow6\mathrm{p}$ resonance. As we will show in the next section, due to the polarizability of the additional valence electron of alkaline-earth atoms  the resulting magic wavelengths are sufficiently far from all resonances such that deep trapping potentials are possible while decay effects due to off-resonant excited-state populations are kept at a sufficiently low level.

\subsection{Decay and loss mechanisms}
As for any optical trap, the small admixture of excited p~states to both the $5\mathrm{s}^2$ ground state and the 5s$n$s Rydberg state leads to a finite photon scattering rate
\begin{equation}\label{eq7}
\gamma_{a}^{b}=\Gamma_{a}^{b}S_{a}^b
\end{equation}
on the respective $a\rightarrow b$ transition for atoms in state $|a\rangle$. Here $\Gamma_a^b$ denotes the rate of radiative decay from level $|b\rangle$ to $|a\rangle$ (cf. figure \ref{fig1}) and the spatially dependent suppression factor
\begin{equation}\label{eq8}
S_{a}^b=\frac{\omega_{\rm L}^2+\omega_{ab}^2}{\left(\omega_{\rm L}^2-\omega_{ab}^2\right)^2}\frac{\wp_{ab}^2E({\bf r})^2}{2\hbar^2}
\end{equation}
corresponds to the admixed fraction of the quickly decaying state $|b\rangle$.

\begin{figure}[t!]
\centering \resizebox{0.75\columnwidth}{!}{\includegraphics{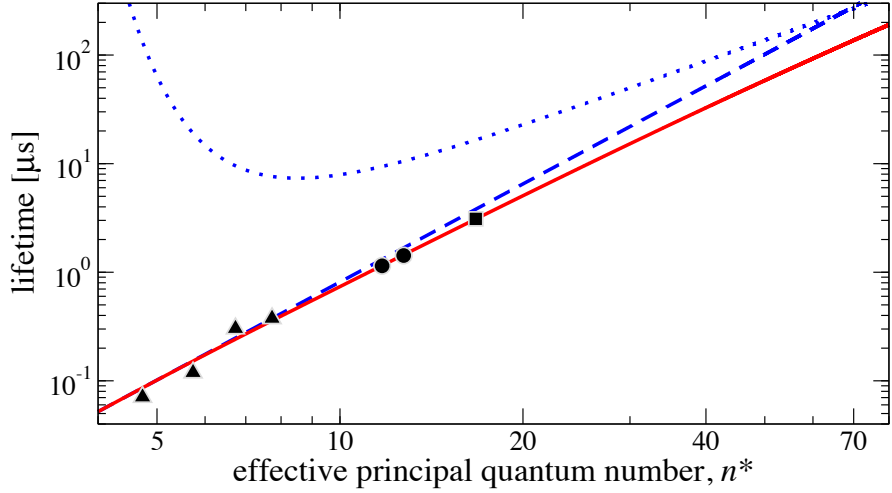}}
\caption{Experimental Sr(5s$n$s) lifetimes from \cite{Gornik1977} (triangles), \cite{Grafstrom1983} (dots), and \cite{Millen2011} (squares), extrapolated to high principal quantum numbers using  (\ref{eq9}) (solid red line). The dashed and dotted blue lines shows the separate contributions of spontaneous decay and black body radiation, respectively.\label{fig3}}
\end{figure}

The excitation to Rydberg states opens up several additional decay channels due to the finite radiative lifetime of the Rydberg state, off-resonant coupling to autoionizing states and direct photoionization of the Rydberg electron. The lifetime of Sr(5s$n$s\,\s) Rydberg states due to spontaneous decay and black body radiation has been measured for low and moderate $n$ in \cite{Gornik1977,Grafstrom1983,Millen2011}. In order to estimate lifetimes of higher excited states we have fitted this data to the expected behaviour at large effective principal quantum numbers $n^{\star}$ \cite{Beterov2009}
\begin{equation}\label{eq9}
\Gamma_{\rm ryd}=\frac{\gamma_{\rm s}}{n^{\star3}}+\frac{\gamma_{\rm bbr}}{n^{\star5}\left(\exp\left(n_{\rm T}^3/n^{\star3}\right)-1\right)},
\end{equation}
yielding $\gamma_{\rm s}=2\times10^8$~Hz, $\gamma_{\rm bbr}=2\times10^9$~Hz and $n_{\rm T}=8.9$. As shown in figure \ref{fig3}, this simple formula yields a rather good fit of the available experimental data.

Besides their spontaneous decay, the 5p$n$s Rydberg states are also unstable against autoionization \cite{Gallagher}. The corresponding rates have been measured over a broad range of principal quantum numbers \cite{Xu1985} and were found to be well described by
\begin{eqnarray}\label{eq10}
\Gamma^{5p_{1/2}}_{\rm ai} &= \frac{6.0 \times 10^{14}}{n^{\star3}}\  {\rm Hz}\;,&\qquad \Gamma^{5p_{3/2}}_{\rm ai} = \frac{9.0 \times 10^{14}}{n^{\star3}}\  {\rm Hz}\;.
\end{eqnarray}

Comparing these expressions to  (\ref{eq9}) we see that autoionization is much more rapid than the spontaneous decay of the 5s$n$s Rydberg state. However, the 5p$n$s states are only weakly mixed into the primary Rydberg state such that its effective autoionization rate 
\begin{equation}\label{eq12}
\gamma_{\rm ai}=\Gamma^{5p_{1/2}}_{\rm ai}S^{5p_{1/2}}_{5s_{1/2}}+\Gamma^{5p_{3/2}}_{\rm ai}S^{5p_{3/2}}_{5s_{1/2}}
\end{equation}
is strongly suppressed.

Finally, the highly excited Rydberg electron may also be lost through direct photoionization  \cite{Saffman2005,Tallant2010,Markert2010}. Photoionization at optical frequencies takes place far above threshold and the corresponding cross sections are expected to be small. The photoionization rate is calculated from \cite{Gallagher}
\begin{equation}\label{eq13}
\gamma_{\rm pi} = E(x)^2\frac{\pi e^2}{4\omega_{\rm L}m_e }\left.\frac{df}{d\varepsilon}\right|_{\varepsilon_{\rm c}}\;,
\end{equation}
where the oscillator strength distribution $df/d\varepsilon$ at the electron's excess energy $\varepsilon_{\rm c}$ has been obtained from a semiclassical expression for the required bound-free dipole matrix elements \cite{Yachkov94}, using the quantum defects from \cite{Beigang1982}. 

\begin{figure}[t!]
\centering \resizebox{0.75\columnwidth}{!}{\includegraphics{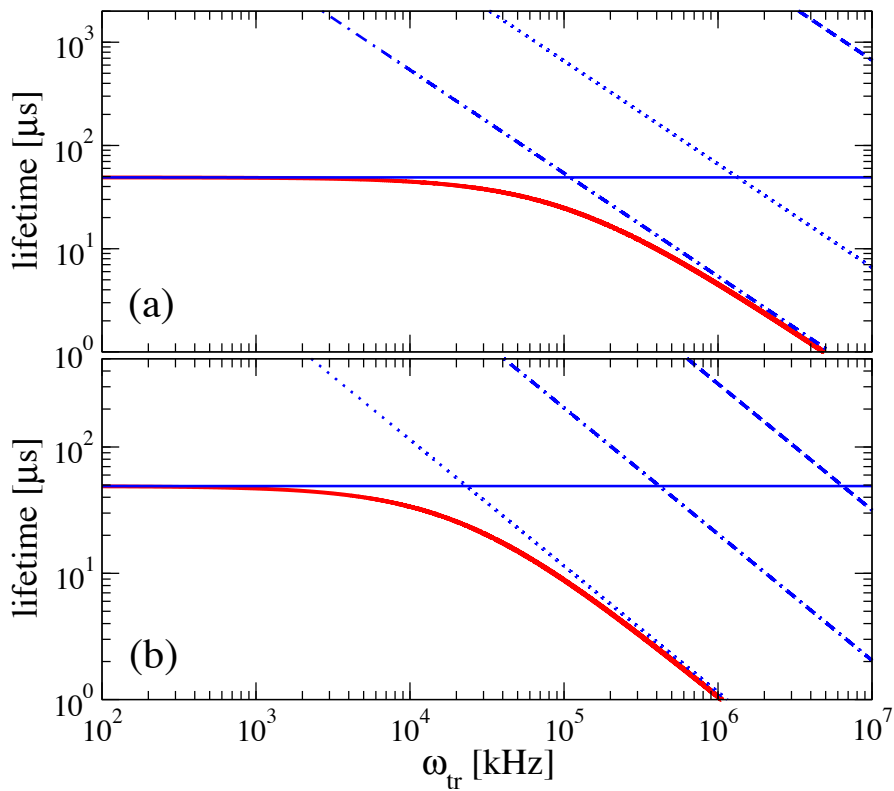}}
\caption{Lifetime of the trapped Sr(5s$^2$\s) ground state and Sr(5s50s\,$^1\mathrm{S}_0$) Rydberg atoms as a function of the local trapping frequency at (a) $\lambda_{\rm m}=323.4$~nm  and (b) $\lambda_{\rm m}=418.6$~nm . The thick red solid line shows the total lifetime of the atoms obtained by summing all decay rates for both ground and Rydberg states. The thin blue lines show the individual contributions from radiative Rydberg states decay (solid line), spontaneous decay of the admixed $p$-states (dashed line), autoionization (dotted line) and photoionization (dash-dotted line).\label{fig4}}
\end{figure}

In order to balance the strong Rydberg-Rydberg atom interactions the optical lattice needs to have large lattice spacings of a few $\mu$m and rather large trapping frequencies of a few $100$~kHz. For, e.g., $a=3\mu$m and $\omega_{\rm tr}=300$~kHz this requires large peak intensities of $2\times10^5$W/cm$^2$ at the maxima of the periodic intensity pattern. At such high intensities, the  intensity dependent loss rates (\ref{eq8}) and (\ref{eq12}) and in particular  the photoionization rate (\ref{eq13}) may become rather large \cite{potvliege2006}. Note, however, that the scalar polarizabilities at the magic wavelengths  (\ref{eq6}) and (\ref{eq7}) are positive. Consequently, atoms in both states are trapped at the local minima of the optical lattice potential, such that all intensity dependent loss mechanisms will be greatly suppressed. Moreover, for tight confinement, the resulting average loss rate can be expressed as a sole function of the local trapping frequency $\omega_{\rm tr}$ and without an explicit dependence on both the peak intensity and the lattice spacing. In this tight-confinement limit we can use (\ref{eq1}) to re-express the electric field amplitude at a local lattice site
\begin{equation}\label{eq14}
E({\bf r})^2=\frac{2U({\bf r})}{\alpha_{\rm m}}=\frac{M}{\alpha_{\rm m}}\omega_{\rm tr}^2 r^2
\end{equation}
in terms of the trap frequency $\omega_{\rm tr}$, where $\alpha_{\rm m}$ is the total polarizability at a magic wavelength ( (\ref{eq5}) and (\ref{eq6})) and $M$ is the mass of the atoms. Assuming that all atoms reside in the lowest vibrational  state $|0\rangle$ at the respective lattice sites, one obtains a simple relation 
\begin{equation}\label{eq15}
\bar{S}^a_b=\frac{3\alpha^a_b}{4\alpha_{\rm m}}\frac{\omega_{\rm L}^2+\omega_{ab}^2}{\omega_{\rm L}^2-\omega_{ab}^2}\frac{\omega_{\rm tr}}{\omega_{ab}}
\end{equation}
between the average suppression factor $\bar{S}^b_a=\langle0|S^b_a|0\rangle$ [cf. equation (\ref{eq8})] and the trap frequency. Likewise the average photoionization rate can be obtained from
\begin{equation}\label{eq16}
\gamma_{\rm pi} = \frac{3\pi e^2\hbar\omega_{\rm tr}}{8\alpha_{\rm m}\omega_{\rm L}m_e }\left.\frac{df}{d\varepsilon}\right|_{\varepsilon_{\rm c}}\;,
\end{equation}
in terms of the trap frequency $\omega_{\rm tr}$.

In figure \ref{fig4} we show our calculated lifetimes as a function of the local trap frequency for Sr($5\mathrm{s}50\mathrm{s}$\,\s) Rydberg states at the two magic wavelengths  (\ref{eq6}) and (\ref{eq7}). As we see, the relevance of the different decay processes strongly depends on the wavelength of the optical lattice. Most importantly, we find that the atomic lifetime is not affected by the additional trapping fields for typical lattice parameters. Even for very large trap frequencies of $\omega_{\rm tr}\sim1$~MHz the total lifetime at both magic wavelengths is solely limited by the Rydberg state decay.

\section{Rydberg-Rydberg atom interactions}
\begin{figure}[t!]
\centering \resizebox{0.75\columnwidth}{!}{\includegraphics{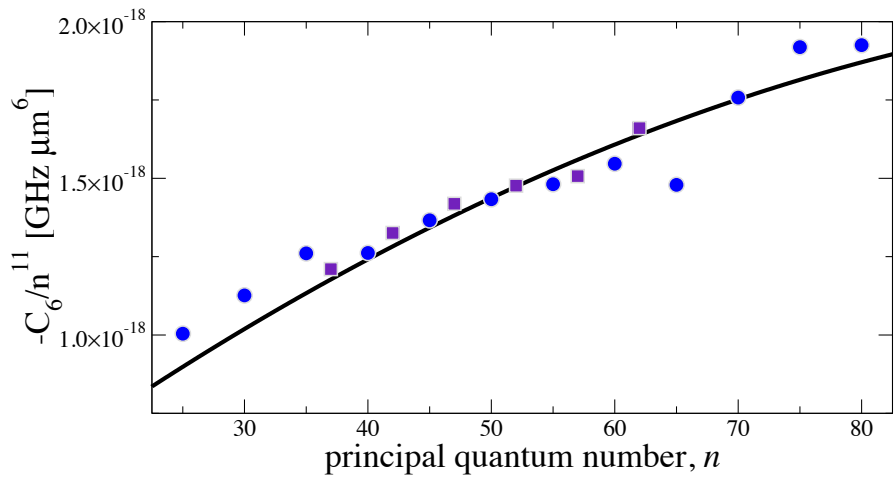}}
\caption{Van der Waals $C_6$ coefficients for the \srs~states, scaled by $n^{11}$. The solid line shows the fit (\ref{eq17}) to our numerical data. The different symbols correspond to the numerical results obtained from two different methods described in the text. \label{fig5}}
\end{figure}

The van der Waals interaction between two 5s$n$s Rydberg atoms arises from off-resonant dipole-dipole coupling to energetically adjacent 5s$n^{\prime}$p-5s$n^{\prime\prime}$p pair states. At high $n$, the 5s$n$s\,\s~Rydberg series is only weakly perturbed \cite{Beigang1982,Esherick1977}, and the relevant dipole matrix elements can be calculated using a single active electron treatment. We calculated two independent sets of  van der Waals  $C_6$ coefficients using wavefunctions obtained from single-electron model-potential calculations \cite{Millen2011}  (circles in figure \ref{fig5}) and from two-electron Hartree-Fock calculations using the effective Sr$^{2+}$ core potential of \cite{greene1991}. As shown in figure \ref{fig4} both approaches yield nearly identical $C_6$ values, which over the depicted range are well described by the simple fit-formula
\begin{equation}\label{eq17}
C_6\,(\mathrm{a.u.})\approx-(1.4+2.2\times10^{-1}n-9.0\times10^{-4}n^2)n^{11}
\end{equation}
For a given $n$ the magnitude of the $C_6$ coefficients is comparable to that for Rb(ns) states. However, in contrast to the alkalis \cite{Walker2008, Singer2005,Reinhard2007}, the van der Waals interaction between \srs~atoms is {\it attractive} for all $n$. As will be shown below, this sign change leads to  profoundly different many-body excitation dynamics.

\section{Many body dynamics}
Having established optimal conditions for our interacting Rydberg atom lattice, we finally give a brief discussion of the resulting many-body dynamics. 
For simplicity, we will restrict the discussion to atoms arranged on a one-dimensional chain. Moreover, we consider a lattice of $N$ tightly confined atoms with negligible spatial overlap between the Wannier states, such that tunneling between lattice sites can be neglected and the quantum statistics of the atoms is not of relevance. The Hamiltonian that governs the dynamics of this system may be split into three parts
\begin{equation}\label{eq18}
\hat{H}=\hat{H}_{\rm L}+\hat{H}_{\rm CoM}+\hat{H}_{\rm I}\;.
\end{equation}
Here, 
 \begin{equation}\label{eq19}
\hat{H}_{\rm L} = \frac{\hbar\Omega}{2}\sum^N_{i=1}\left(\hat{\sigma}_{eg}^{(i)}+\hat{\sigma}_{ge}^{(i)}\right)-\hbar\Delta\sum^N_{i=1}\hat{\sigma}_{ee}^{(i)}
\end{equation}
describes the coupling of the Rydberg laser, where $\Omega$ and $\Delta$ are the respective Rabi frequency and detuning (cf. figure \ref{fig1}), and the operators $\hat{\sigma}_{\alpha\beta}^{(i)}=|\alpha_i\rangle\langle\beta_i|$ denote atomic transition and projection operators for the dressed ground ($|g_i\rangle$) and Rydberg state ($|e_i\rangle$) of the $i$th atom. Introducing relative atomic coordinates ${\bf r}_i$ with respect to the corresponding sites ${\bf a}_i=i\cdot a {\bf e}_x$ of the atomic chain (aligned along the x-axis), the Hamiltonian of the atomic center-of-mass (CoM) dynamics
\begin{equation}\label{eq20}
\hat{H}_{\rm CoM}=\sum_{i=1}^{N}\frac{\hat{{\bf p}}^2_i}{2M}+\sum_{i=1}^N\frac{1}{2}M\omega_{\rm tr}^2r_i^2
\end{equation}
is a simple sum of harmonic oscillators. The atomic motion and the internal state dynamics are coupled by the interaction Hamiltonian
\begin{equation}\label{eq21}
\hat{H}_{\rm I}=\sum_{i<j}\frac{C_6}{\left|{\bf a}_i+{\bf r}_i-{\bf a}_j-{\bf r}_j\right|^6}\hat{\sigma}_{\rm ee}^{(i)}\hat{\sigma}_{\rm ee}^{(j)}\;.
\end{equation}
If the trapping potential is not identical for the ground and Rydberg states, the interactions will lead to entanglement between the atomic internal and CoM degrees of freedom and may ultimately cause heating of the atoms. In the following, we will show that for sufficiently strong confinement this interaction-induced coupling is adiabatically eliminated, giving rise to an effective lattice Hamiltonian for spatially frozen spins.

\subsection{Atomic motion}
Since for tight confinement the extent $\sigma=\sqrt{\hbar/M\omega_{\rm tr}}$ of the local atomic CoM wavefunctions is much smaller than the lattice spacing $a$, we can expand the van der Waals potential in  ($\ref{eq21}$) 
\begin{equation}
\frac{C_6}{\left|{\bf a}_i+{\bf r}_i-{\bf a}_j-{\bf r}_j\right|^6}\approx \frac{C_6}{a^6(i-j)^6}\left[1-6\frac{x_i-x_j}{a(i-j)}\right]
\end{equation}
to leading order in the atomic CoM displacements ${\bf r}_i$ and rewrite the CoM Hamiltonian 
\begin{eqnarray}\label{eq22}
\hat{H}_{\rm CoM}+\hat{H}_{\rm I}&=&V_0\sum_{i<j}\frac{\hat{\sigma}_{\rm ee}^{(i)}\hat{\sigma}_{\rm ee}^{(j)}}{(i-j)^6}+\sum_i\hbar\omega_{\rm tr}\hat{a}^{\dagger}_i\hat{a}_i\nonumber\\
&&+\tilde{V}_0\sum_{i<j}\left(\hat{a}^{\dagger}_j-\hat{a}^{\dagger}_i+\hat{a}_j-\hat{a}_i\right)\frac{\hat{\sigma}_{\rm ee}^{(i)}\hat{\sigma}_{\rm ee}^{(j)}}{(i-j)^7}
\end{eqnarray}
in terms of creation and annihilation operators, $\hat{a}^{\dagger}_i$ and $\hat{a}_i$, of vibrational states along the chain axis at a given site $i$. Here $V_{0}=C_6/a^6$ is the nearest neighbor interaction between adjacent sites. We see that the Rydberg interactions not only yield the desired level shifts of excited pair states, but also lead to intra-band coupling with a coupling strength $\tilde{V}_0=6\sigma V_0/(a\sqrt{2})$. This coupling can, however, be strongly suppressed by realizing sufficiently strong trapping potentials, for which the  vibrational splitting $\hbar\omega_{\rm tr}$ exceeds the coupling strength $\tilde{V}_0$. We can estimate its effect by assuming $\tilde{V}_0<\hbar\omega_{\rm tr}$ to derive an effective single-band Hamiltonian for the lowest band of the atomic lattice. In second order perturbation theory in $\tilde{V}_0/\hbar\omega_{\rm tr}$ one obtains
\begin{equation}\label{eq23}
\hat{H}_{\rm CoM}+\hat{H}_{\rm I}=V_0\sum_{i<j}\frac{\hat{\sigma}_{\rm ee}^{(i)}\hat{\sigma}_{\rm ee}^{(j)}}{(i-j)^6}\left[1+\eta_{ij}\right]\;,
\end{equation}
which shows that the motional intra-band coupling simply yields a correction $\eta_{ij}=\frac{V_0}{\hbar\omega_{\rm tr}}\left(\frac{6\sigma}{a}\right)^2(i-j)^{-1}$ to the van der Waals shifts.

To be specific let us consider our preceding example of $\omega_{\rm tr}=300$~kHz and $a=3\mu$m. For a typical Rydberg state with $n=50$ the resulting correction factor $\eta_{ii+1}=0.05$ is indeed negligibly small. This simple discussion shows that, for reasonable experimental parameters, atomic motion can be practically frozen out, despite the presence of strong interactions ($V_0=9.6$~MHz) that greatly exceed the energy scale of the lattice confinement. Such large interactions are essential to assure a sufficiently short timescale for the many body dynamics of the internal states, which must be faster than the decay of the Rydberg lattice.

\begin{figure}[t!]
\centering \resizebox{0.75\columnwidth}{!}{\includegraphics{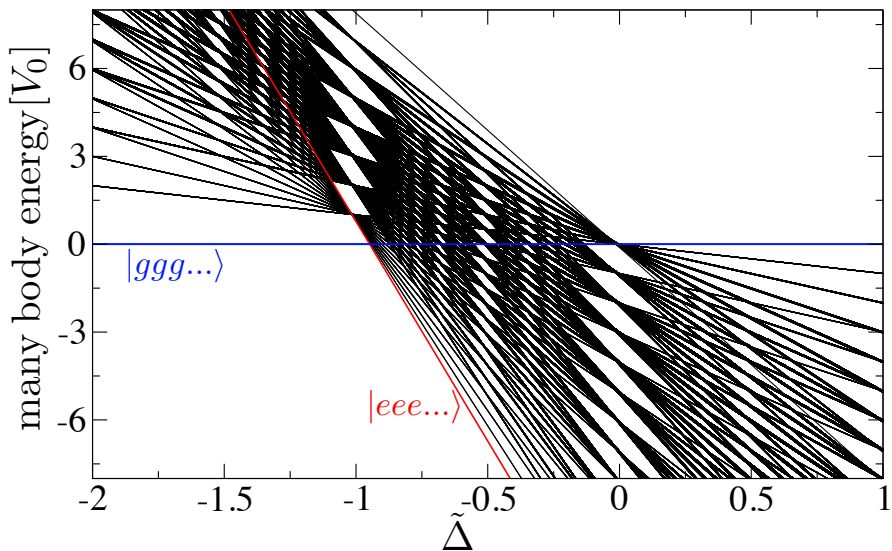}}
\caption{Many-body energy spectrum of an \emph{attractively} interacting Rydberg atom chain of $N=15$ atoms for $\tilde{\Omega}=0$, showing a single ground state crossing between the many-body ground state $|G\rangle\equiv|ggg...\rangle$ (blue line) and the fully excited state $|E\rangle\equiv|eee...\rangle$ (red line). \label{fig6}}
\end{figure}

\subsection{Excitation dynamics}
\begin{figure}[t!]
\centering \resizebox{0.75\columnwidth}{!}{\includegraphics{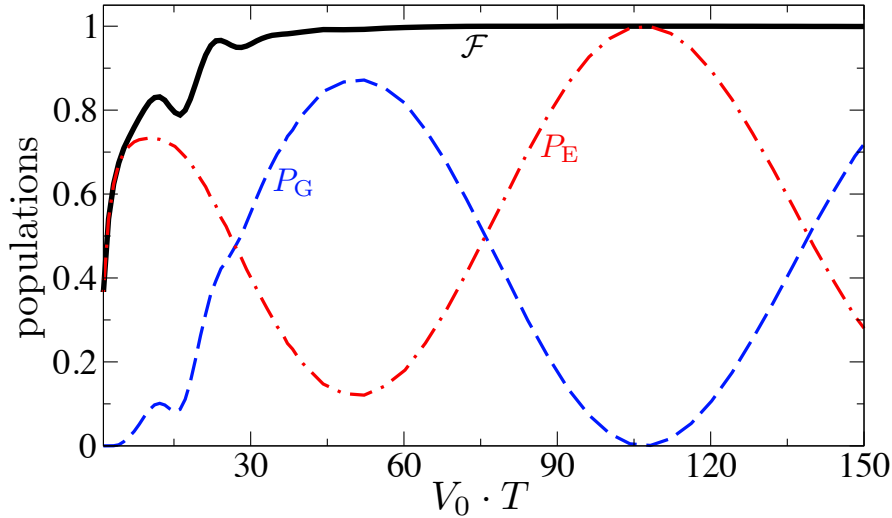}}
\caption{Final populations $P_G$ (blue dashed line), $P_E$ (red dot-dashed line ) of the many-body ground state $|G\rangle$ and the fully excited state $|E\rangle$ respectively,  as a function of the duration of the Rydberg excitation pulse for a chain of $N=15$ atoms. The thick black line shows the fidelity ${\mathcal{F}}=P_{\rm E}+P_{\rm G}$.\label{fig7}}
\end{figure}

Following the preceding discussion, we start from our reduced Hamiltonian
\begin{equation}\label{eq24}
\hat{H}=\frac{\tilde{\Omega}}{2}\sum_i\left(\hat{\sigma}_{eg}^{(i)}+\hat{\sigma}_{ge}^{(i)}\right)-\tilde{\Delta}\sum_i\hat{\sigma}_{ee}^{(i)}-\sum_{i<j}\frac{\hat{\sigma}_{\rm ee}^{(i)}\hat{\sigma}_{\rm ee}^{(j)}}{(i-j)^6}
\end{equation} 
where we have introduced the dimensionless Rabi frequency $\tilde{\Omega}=\hbar\Omega/|V_0|$ and laser detuning $\tilde{\Delta}=\hbar\Delta/|V_0|$. Lattice Hamiltonians of this type have recently been investigated by several authors \cite{Pohl2010,Schachenmayer2010,Weimer2010,Lesanovsky2011}. While previous work has focussed on repulsive Rydberg-Rydberg atom interactions, the present case of attractive interactions ($C_6<0$) leads to qualitatively different behaviour. Consider, for example, the many-body ground state phase diagram of the Hamiltonian  (\ref{eq24}), spanned by $\tilde{\Omega}$ and $\tilde{\Delta}$. Here, it was found that repulsive interactions cause a series of energy crossings between different Rydberg lattice states as a function of $\tilde{\Delta}$, which leads to the formation of Rydberg atom crystals upon adiabatic excitation with chirped laser pulses \cite{Pohl2010}. For attractive interactions, on the other hand, this series is replaced by a single curve crossing between the $N$-atom ground state $|G\rangle\equiv|gggg...\rangle$ and the fully excited many-body state $|E\rangle\equiv|eeee...\rangle$ (cf. figure \ref{fig6}). In the classical limit $\tilde{\Omega}=0$ we find a real crossing at 
\begin{equation}\label{eq25}
\tilde{\Delta}_{\rm c}=-N^{-1}\sum_{i<j}\frac{1}{(i-j)^6}\;,
\end{equation}
which turns into an avoided crossing at finite $\tilde{\Omega}$. Upon dynamic changes of $\tilde{\Omega}$, i.e. pulsed Rydberg excitation, this can be exploited to create coherent superposition states between $|G\rangle$ and $|E\rangle$. To illustrate the physical mechanism, we consider the most simple scenario of an excitation pulse at a fixed laser frequency. Figure \ref{fig7} shows the resulting excitation dynamics, obtained by exact diagonalization for $N=15$ atoms. The laser is slightly red detuned from $\tilde{\Delta}_{\rm c}$ such that the initial state with all atoms in their individual ground states coincides with the initial many body ground state. The initial increase of $\tilde{\Omega}$ lifts the near-degeneracy of $|G\rangle$ and $|E\rangle$, and, more importantly, leads to Landau-Zener transitions between these two many-body states at the wings of the excitation pulse. Consequently, at the end of the pulse, one obtains a collective excitation of $N$-atom superpositions of $|G\rangle$ and $|E\rangle$, whose composition is controlled by the pulse length $T$. In particular, we find a high fidelity $N$-atom GHZ state, for a rather short pulse length of $V_0\cdot T\approx27$ (cf. figure \ref{fig7}b). For the above parameters, this corresponds to a preparation time of $0.4\ \mu$s, which is much smaller than the total atomic lifetime calculated above.

\section{Conclusion and Outlook}
In summary, we have given a detailed analysis of optical trapping of alkaline-earth Rydberg atoms, that exploits the availability of two valence electrons.
This work highlights the great potential of alkaline-earth atoms to provide new ways of studying and manipulating cold Rydberg gases. We have demonstrated that the polarizability of the additional valence electron can be used to create {\emph{deep}} high-field seeking or low-field seeking optical potentials for Rydberg atoms. Focussing on the particular case of strontium atoms, we have identified accessible magic wavelengths that permit the simultaneous confinement of ground and Rydberg states in identical trapping potentials. Such a trap provides an ideal setting for studying many-body spin dynamics, as the interactions no longer couple the internal and external states if the confinement at each lattice site is sufficiently strong. 
We show that even at rather high intensities, as required for strong confinement, the total lifetime of the atomic lattice is limited only by spontaneous decay of the Rydberg state, which is essential for its applicability to quantum simulation and quantum information processing schemes.
In contrast to the alkalis, the van der Waals interactions for the \srs~Rydberg series are found to be attractive. As we have shown, the many-body dynamics of a Rydberg atom lattice  is qualitatively different for attractive, rather than repulsive, interactions. In particular, they can be used to prepare highly-entangled $N$-atom GHZ states in a single excitation step by appropriately chosen pulses. The applicability of such Rydberg-based schemes to high-precision frequency metrology in strontium optical lattices may be subject of future work. While we have focussed on \s~Rydberg states, higher angular momentum states could also be trapped in an analogous way.

Attractive interactions between non-degenerate Rydberg states are also appealing fin the context of recently proposed techniques for engineering effective ground state atom interactions by off-resonant Rydberg dressing of Bose-Einstein condenstates \cite{Henkel2010,Pupillo2010,Cinti2010}. Quantum degeneracy was recently achieved for strontium atoms \cite{Stellmer2009,Escobar2009,Salvo2010}. In contrast to alkali condensates, here Rydberg dressing using the Sr(5s$n$s) states could be used to realize attractive effective ground state atom interactions, which, for example, may permit the creation of three-dimensional bright solitons \cite{Maucher2011}.

The discussion of the internal state dynamics was based on the most simple setting of a single ground state coupled to one Rydberg state. Recently, there have been several theoretical proposals describing new schemes for quantum simulations \cite{Gorshkov2010,Hermele2009} and quantum information processing \cite{Daley2008,Gorshkov2009}, based on the particular level structure of low-lying states in two-electron atoms. We anticipate that the combination of these more sophisticated coupling schemes, specific to alkaline earth atoms, with the availability of strong and long-range Rydberg-Rydberg atom interactions will offer new perspectives for such applications.

\ack{M.P.A.J. acknowledges discussions with P.~Pillet on the possibility of trapping using the second electron. J.M. and M.P.A.J. also thank J.~Pritchard, C.~S.~Adams and R~M.~Potvliege for useful discussions.}

\end{document}